\newcommand{\beqn}{\begin{equation}}
\newcommand{\eeqn}{\end{equation}}
\newcommand{\beqa}{\begin{eqnarray}}
\newcommand{\eeqa}{\end{eqnarray}}
   \let \eps=\epsilon 
\let\al =\alpha 

\documentstyle[11pt]{article}
\title{Ab-initio-calculations of the GMR-effect in Fe/V
multilayers\thanks{based on the diploma thesis of A.~Moser,
 Regensburg 1997 (present address: Siemens Co., Munich)
 ;\,\,$^1$ corresponding author; e-mail:
krey@rphs1.physik.uni-regensburg.de;
 $^2$ present address: Inst. f\"ur Physikalische Chemie der
LM-Universit\"at M\"unchen } }

 \author{A. Moser, U. Krey$^1$,
A. Paintner$^2$, B.
Zellermann$^2$
\\  
Institut f\"ur Physik II der Universit\"at Regensburg,\\
Universit\"atsstr.~31, D-93040 F.R.G. } 
       
\begin{document}

\large
\date{October 7, 1997; revised version}
\maketitle
\hrule
\begin{abstract} 
\noindent 
In a self-consistent semi-empirical numerical approach based on {\it
ab-initio}-calculations for small samples,
 we evaluate the GMR effect for disordered
(001)-(3--Fe/3--V)$_\infty$ multilayers by means of  a   Kubo formalism.
We consider four different types of disorder arrangements: In case (i)
and (ii), the disorder consists in the random interchange of some Fe
and V atoms, respectively, at interface layers; in case (iii) in the
formation of small groups of three substitutional Fe atoms in a V
interface layer and a similar V group in a Fe layer at a different
interface; and  for case (iv) in the substitution of some V atoms in the
innermost V layers by Fe. For cases (i) and (ii), depending on the
distribution of the impurities, the GMR effect is enhanced or reduced by
increasing disorder, in case (iii) the GMR effect is highest, whereas
finally, in case (iv), a negative GMR is obtained ( ''inverse GMR'').
\end{abstract}

magnetotransport effects (metals/alloys);

\section{Introduction}
The so-called Giant Magnetoresistance effect (GMR-effect) is well-known
meanwhile, \cite{l:Baibich}, and several overviews exist on  theoretical
interpretations, \cite{l:Levy,l:Mertig,l:Zahn}. The effect exists in
 thin film systems made of metallic ferromagnets separated by
nonmagnetic or antiferromagnetic metallic spacers, and consists in the
fact that e.g.\ in such a trilayer or multilayer system, if one starts
from a state, where the ferromagnetic films have mutually antiparallel
magnetization,  one can switch the
magnetization directions,  by application of a magnetic field, to
parallel orientation, which -- as a consequence -- generally implies a
decrease of the electrical resistance. This amounts typically to values
around 10\% or so, which is already interesting for applications, but in
the case of Fe/Cr/Fe trilayers the effect can be as large as 200\%,
\cite{l:Gruenberg}.

Thus, the GMR-effect is given by the simple formula

\beqn \label{e:Gmr}
{\rm GMR}
= \frac{\Gamma_{\uparrow \uparrow}} {\Gamma_{\uparrow \downarrow}}-1 \,,
\eeqn
where the $\Gamma_{\uparrow \uparrow}$ and $\Gamma_{\uparrow \downarrow }$
are the conductances  for mutually parallel resp.~antiparallel
 magnetizations. Here one distinguishes between the CPP
 and CIP geometries, where the current is {\it perpendicular} to the
 planes, or {\it in} the plane, respectively.

Considering Eqn.~(\ref{e:Gmr}), we would like to stress already at this
place that  in a numerical calculation not $\Gamma_{\uparrow \uparrow}$
and $\Gamma_{\uparrow \downarrow}$ separately, but only their {\it
ratio} $\Gamma_{\uparrow \uparrow}/\Gamma_{\uparrow \downarrow}$ must
come out correctly. 

A considerable GMR effect can already exist in the ballistic regime,
i.e.~without any impurities, due to the fact that the reflectivity of the
electrons at the interfaces changes with the above-mentioned switch.
This is shown by the {\it ab initio} calculation of Schep {\it et
al.} for Co/Cu-multilayers in \cite{l:Schep}, and by  recent model
calculations of Krompiews\-ki {\it et al.},
\cite{l:Krompiewski,l:Krompiewski1}. But one of the main problems,
namely the question, how impurities and disorder influence the strength
and perhaps even the sign of the GMR in a {\it realistic} system with
non-ideal interfaces and significant impurity scattering, is not
answered by theories for the ballistic case. For example, it seems
almost natural to state that an increase of spin-dependent scattering at
the interface should  lead to an {\it enhancement} of the GMR; on the
other hand one can also imagine that too much disorder at the interface
should {\it reduce} the difference of the  spin-dependent reflexion
properties for the two cases considered above: Thus, without detailed
calculations for different systems it remains an open question, whether
the GMR is enhanced or reduced by an increasing amount of impurities.
Moreover, the answer on this question  may depend on the {\it
arrangement} of the impurities.

In the present communication, we study the
influence of disorder on the GMR for bcc-(001)-(Fe-3/V-3)$_\infty$ multilayers,
both for CPP and CIP geometry, with four different situations (see below).
 Our extensive numerical 
 calculations employ a self-consistent semi-empirical approach,
which also works for {\it non-ideal} systems with impurities. We have applied
a similar approach already earlier for {\it
almost-ab-initio}-calculations of the magnetic and transport properties
of strongly disordered or even amorphous systems. Both the electronic
structure, \cite{l:Krey1,l:Krey2,l:Paintner}, and the transport
properties of the systems, \cite{l:Paintner,l:Krey3}, have been
calculated, the transport properties directly from the Kubo formula, and
separated into contributions from the up-spin and down-spin carriers. We
stress that the Kubo approach is rigorous in principle, and  does not
invoke the usual Boltzmann approximations. Instead, already from the
beginning the disorder of the system is  fully taken into account:
In particular, in our paper the eigenvalues and eigenvectors of the
electronic Hamiltonian are always calculated for the {\it disordered}
system, before the Kubo formula is applied (see below); in the language
of diagrammatic theories this means that the vertex corrections are
automatically included.

However, all this is possible only for rather
 small systems: Our computer samples comprise stacks of 12
 non-equivalent, partially disordered  Fe-- or V--monolayers with
 4$\times$4 atoms per layer, and with periodic boundary conditions in
 all three directions. Thus, altogether we have 192 atoms, with 9
 orbitals per atom (five 3d--, three 4p--, and one 4s--orbital);
 i.e.~for fixed spin projection $s=\pm 1$ of the electron we diagonalize
 a Hamiltonian with 1728 lines and columns. After each diagonalization,
 the expectation values of the local occupations and local moments for
 every orbital are calculated, and the Hamiltonians are updated, until
 finally, after a lot of iterations of the procedure, self-consistency
 is obtained with respect to all local charge and spin expectations (see
 below). Also the Fermi energies $E_f(s)$ are determined
 self-consistently. However, in view of the smallness of our systems,
 the results suffer from  the fact that in the numerical calculation we
 have a discrete spectrum and not a continuum, so that evaluation of
 histograms instead of continuous functions becomes necessary. (This is
 different with the technique of Asano {\it et al.}, \cite{l:Asano},
 which is however not applicable to our system, see below.)
 Furthermore, one should also be aware of the fact that the T=0
 conductance of a mesoscopic physical system is not self-averaging: One
 needs to average over different samples to obtain significant ''typical
 results'', although with large error bars. But even for these small
 systems, such results can be obtained, as seen below. This has also
 been exemplified in former calculations of the resistivity in the
 magnetic state of disordered systems, \cite{l:Paintner,l:Krey3}.

 In the following sections we describe at first our formalism, and then
our results for (001)-(3--Fe/3--V)$_\infty$--multilayers with four
different types of disorder: In cases (i) and (ii), we consider {\it
interchanges} of some nearest--neighbour Fe and V atoms at the
interface, whereas in the particular  case (iii)  we deal with the
effect of  small Fe "islands" of three Fe atoms in a V interface layer
and {\it vice versa}; finally, the case (iv) of a substitutional Fe
impurity in an {\it inner} V layer is considered.

In case (i), we treat the standard situation that
the impurities are randomly distributed among all four nonequivalent
interfaces, whereas in case (ii) we assume that  the impurities
are concentrated at only one of the interfaces.
Concerning (i), we find at first -- as expected -- an {\it increase}
 of the GMR with increasing impurity concentration, namely from
GMR$\sim$ 60\%  for an impurity concentration of
$n_{\rm {\scriptsize{imp}}}\approx 1$\%
  to GMR$\sim$ 120\% for $n_{\rm{\scriptsize{imp}}}\approx 3$\% and
 $\approx 4$\%, whereas in case (ii)
a {\it reduction} of the GMR from $\sim 60$\%
 (for $n_{\rm {\scriptsize{imp}}}\approx 1$\%)
 to $\sim$ 0\% (for $n_{\rm {\scriptsize{imp}}}\approx $5\%) is obtained.
Also in case (i) the GMR decreases again, if
$n_{\rm {\scriptsize{imp}}}$ becomes as large as
$\approx 5$\%. This different behaviour is discussed below.

For the  particular  case (iii), see above, we get
the largest values of the GMR ($\sim$ 250 to 300\%). Finally, in case
(iv), the Fe impurity in the inner V layer is significantly magnetized
by --0.5 $\mu_B$, i.e.~antiparallel to the Fe layers, if these are
aligned in parallel (see below); whereas the Fe impurity is nonmagnetic,
if the magnetization directions of the Fe films are mutually antiparallel. In
the first-mentioned case the impurity scattering is stronger, which
implies that in case (iv) we have a negative GMR (''inverse GMR'') of
$\sim$ (--50\%). This is true for both  the CPP and the CIP geometries,
although quantitatively the results are somewhat different for these
geometries (see below).

A negative GMR has already been observed in \cite{l:Mertig2}, but for
{ternary} systems, whereas the present system is {\it binary}.

\section{Formalism}
 We use our realistic self-consistent
  semi-empirical LCAO approach already described in
 \cite{l:Suess,l:Paintner}, i.e.\ with 9 orthogonalized orbitals per
 atom (five 3d-, three 4p- and one 4s-orbital) and a collinear magnetic
 state described by the equations
  \beqa  \label{e:Hamiltonian}
 \sum_{m\beta} H_{l\alpha ,m\beta}\, c_{m\beta ,s}^{(\nu)}
 +\frac{U_{l\al}}{2}\,\langle n_{l\al\uparrow}+n_{l\al\downarrow}-
 2\,n_{l\al}^{para}\rangle\,c_{l\al,s}^{(\nu)}\nonumber\\
 -\frac{U_{l\al}}{2}\,\langle n_{l\al\uparrow}-n_{l\al\downarrow}
 \,\rangle\cdot s\cdot c_{l\al,s}^{(\nu)} =\epsilon_\nu (s)\cdot
 c_{l\al,s}^{(\nu)}\,. \eeqa
 In Equ.~(\ref{e:Hamiltonian}) the $c_{l\al,s}^{(\nu)}$ are the
 probability amplitudes for the event that an electron with spin $s$ (
 $=\pm 1)$ and single-particle energy $\eps_\nu (s)$ occupies an orbital
 $\al$ $(=1,...,9)$ at the site $l$. Here $\nu =1,...,N_l\cdot N_\al$,
 where the number of atoms is $N_l=192$ and the number of orbitals per
 atom  $N_\al=9$. The orbitals are assumed to be orthonormalized, and
 the real-symmetric matrix $H_{l\al ,m\beta}$ describes the paramagnetic
 state of the disordered system.
 Altogether 26 neighbours, i.e.\ up to the third shell, are taken into
 account for each site. The matrix elements are derived from 
 Papaconstantopoulos, \cite{l:Papa}, in the approximation with
 two-center integrals: We  assume that we have a bcc-structure with
 an averaged lattice constant, $a=(a_{Fe}+a_V)/2$, with $a_{Fe}=2.87$
 $\AA$ and $a_V=3.02 $ $\AA$. Thus it is only necessary to modify
  the hopping matrices of Papaconstantopoulos 
  according to the modified positions, i.e.~for the two-center integrals
  one uses relations as $I_{dd}(r)\propto r^{-5}$, see
  \cite{l:Suess,l:Paintner}; additionally, if site $l$ is occupied by a
  Fe atom, but site $m$ by V, we assume as usual $H_{l\al
  ,m\beta}^{Fe,V}=(H_{l\al,m\beta}^{Fe,Fe}\cdot
  H_{l\al,m\beta}^{V,V})^{1/2}$. Finally, to get agreement with {\it ab
  initio} calculations, see below, we have used a  common shift of
  $\Delta E = 0.5$ eV for the Fe d-levels with respect to the values in
  \cite{l:Papa}, whereas for V no shift was assumed, and for the
  Hubbard-energies we have taken $U_{l\al}=5.8$ and $2.58$ eV for the Fe
  and V d-orbitals, respectively, $U_{l,\al}=0$ otherwise.

Concerning the expectation values in Equ.~(\ref{e:Hamiltonian}), we
require {\it self-consistency} in the magnetic state, again for the
disordered system, namely for  every site
and every d-orbital we demand that
    $ \langle n_{l\al ,s}\rangle \equiv
     \sum_{\nu =1}^{\nu_f(s)}\,|c_{l\al,s}^{(\nu )}|^2$.
Here $\nu_f(s)$ counts the highest occupied single-particle eigenstate
for $s=\pm 1$, respectively.

In Fig.~1, the results for the magnetic moments of the different layers for
an ideal sample of our system are presented, in comparison with similar
results obtained by us for the same system with an {\it ab-initio} LMTO
method, \cite{l:Anderson}. Obviously, the agreement obtained is quite
convincing and should give confidence to the reliability of our
self-consistent semi-empirical method. Moreover, the following results from
Fig.~1 deserve attention: In the central Fe layers, the Fe moments are
enhanced to 2.8 $\mu_B$ with respect to the bulk value of 2.2 $\mu_B$,
whereas at the interface they are reduced to 1.6 $\mu_B$. On the other hand,
Vanadium, which is nonmagnetic in the bulk, has at the interface layers a
moment of $(-0.5
\,\mu_B)$, i.e.\ antiferromagnetically coupled to Fe. Already the second
V layer, however, is practically nonmagnetic. These results are similar
to those obtained by the first-principles LMTO calculations of
 F.\ S\"uss, \cite{l:SuessPoz,l:SuessDiss}.

\section{The Kubo formalism}
 The resistivity is calculated by means of
the Kubo formula, \cite{l:Mott}, namely
\beqn  \label{e:Kubo}
\sigma_{xx}^{(s)} = \frac{e^2\pi}{\hbar\Omega}\,
\overline {(g^{(s)}(E_f))^2\,\,
(\Delta E(s))^2\,|\langle f,s|x|i,s\rangle |^2}\,. \eeqn
Here $\Omega$ is the volume of the elementary cell of 192 atoms,
$\sigma_{xx}^{(s)}$ the contribution of electrons with spin direction
$s=\pm 1$ to the conductivity in x-direction, $g^{(s)}(E_f)$ the value of the
spin-dependent density of states of the disordered system
 at the Fermi energy $E_f(s)$; $|i,s\rangle$ and
$|f,s\rangle$ are the exact eigenstates of the full Hamiltonian, again
with impurities, with single-particle eigenvalues
just above above and below $E_f(s)$, respectively, see below;
 $\langle ...\rangle$ denotes the quantum
mechanical expectation value, and the overline denotes an average over
10 samples. $\Delta E(s)$ is a typical energy difference
involved in the transition from $i$ to $f$. To be precise: We choose
$\Delta E(s) =\eps^{\nu_f+1}(s)-\eps^{\nu_f-1}(s)$, if the highest occupied
single-particle eigenstate of spin $s$ has single-particle energy
$\eps^{\nu_f}(s)$. Furthermore:
 although the eigenstates of our Hamiltonian have
been calculated with periodic boundary conditions in x, y, and
z-direction, we assume in Eqn.~(\ref{e:Kubo}) for the calculation of the
resistivity that the resistance is measured with parallel
planar contacts of a
distance as small as our cluster-sizes
 $\Delta x=\Delta y=4\,a=11.78$ $\AA$ and
$\Delta z=12\cdot (a/2)=17.67$ $\AA$, respectively. Thus, even without
impurities we assume an inelastic dephasing-length of this short size,
i.e.~by a factor (2/3) shorter in x- and y-directions than for the
z-direction. As a consequence of this factor (2/3), our in-plane
conductivities should be scaled by a factor (3/2)$^2$=2.25, if a
direct comparison with the {\it perpendicular}
 conductivity is desired. In
any case, these correction factors do not enter  the GMR, since the
ratio $\Gamma_{\uparrow \uparrow}/\Gamma_{\uparrow\downarrow}$ in
Eqn.~(\ref{e:Gmr}) does not depend on them.

Concerning the average of Eqn.~(\ref{e:Kubo}), for $|i,s\rangle$ and
$|f,s\rangle$  we take the $(n+1)$  highest occupied and $n$ lowest
unoccupied states, respectively, for given $s$, with
$n$=1,...,5  in Fig.~2 below, n=1 otherwise.
 Additionally, as it is usual with the
Kubo-formalism for the dc-conductivity, although it would be rigorous
only in the thermodynamic limit, we also include the case $i$=$f$ in
eqn.\ (\ref{e:Kubo}), so that the average in this equation is dominated
by  the $2n+1$ diagonal terms. Finally, the origin of our coordinates is
fixed in such a way that the matrix element $\langle i,s|x_k
|i,s\rangle$ would give the actual length of our elementary cell in
$k$-direction  ($k=x,y,z$), i.e.~the distance of the contacts,
 for constant $|i,s\rangle$.

\section{Results for the GMR}
\subsection{Pure sample, cases (i) and
(ii), and some remarks}
 In Fig.~2, which only should be considered as a check of the accuracy
of our method and also serves for the statement (see below) that
without impurities we obtain GMR$\approx$0, we present our (fictitious)
results for the CPP- resp.~CIP-conductivities of pure samples obtained
with the different dephasing lengths of $\Delta z=17.67$ $\AA$ and
$\Delta x=\Delta y= 11.78$ $\AA$, corresponding to our sample size, see
above. Taking the average over the five cases of $(2n+1)$ one gets the
dotted lines, from which one concludes
$\rho_{\rm{\scriptsize{CIP}}}/\rho_{{\rm{\scriptsize}CPP}}\sim 2 $,
 as expected from $(\Delta z/\Delta x)^2=2.25$. The error bars of our
 results amount to $\sim \pm $ 20\%. Since these results apply both to
 the cases of mutually antiparallel resp.~parallel  magnetizations of the Fe
 films, we conclude that in our case {\it the GMR-effect in the pure
 systems vanishes within our accuracy}. For the disordered systems this will
be different, see below.

Additionally, at this place the following remarks are in order:
 
1.) It is not important that due to our short dephasing length our values for
$\rho_{\rm{\scriptsize{CPP}}}$ and $\rho_{\rm{\scriptsize{CIP}}}$ are
 $\sim 2$ resp.~$\sim 4$ times larger than the estimates, which one
would expect at room temperature, \cite{l:REMLandolt}:  Since
these rescaling factors of $\sim 2$ resp.~$\sim 4$ do not depend on
whether one considers $\Gamma^{\uparrow\uparrow}$ or
$\Gamma^{\uparrow\downarrow}$, the GMR itself should be insensitive
against these rescalings, as already mentioned in connection with
Eqn.~(\ref{e:Gmr}). 

2.) This is supported by the observation that the numerical results for
the resistivities in the magnetic state of disordered Fe/Ni/Mn alloys
 in  our former paper \cite{l:Paintner} have also  come out 
 too large just by a constant factor $\sim 5$, in spite of the fact that
with increasing Mn concentration the conductance decreased considerably.
However again, apart from the constant factor, the {\it concentration
dependence} of the experimental values was well-reproduced in
\cite{l:Paintner} with our formalism. 

3.) Y.~Asano {\it et al.}, \cite{l:Asano}, have performed a schematic
 model calculation for pure and impure systems with only s-bands, only
 nearest-neighbour hopping in simple-cubic arrays, and
where the magnetism was not treated self-consistently, but replaced by a
spin-dependent constant potential $V_\uparrow =-0.5$,
$V_\downarrow=+0.5$ for the magnetic atoms, whereas for the nonmagnetic
atoms  $V\equiv +0.5$ was chosen. Finally, even the Fermi energy was
arbitrarily fixed at $E_f=0$. (Here the magnitude of the
nearest-neighbour hopping integral has been used as energy unit.) Due to
these simplifications, Asano {\it et al.} could use a {\it
recursive technique}, where planes with 12$\times$12 atoms could be
treated, and 'perfect leads' could be attached to the  sample in the
current direction. Although within our group this powerful technique
has already been extended to extremely accurate model calculations of
the CPP-GMR and of a corresponding Giant Magneto-Thermopower in pure
s-band tight-binding samples with infinite planes, \cite{l:Krompiewski},
the method is not applicable to the present system, since s-, p-, and
d-bands, self-consistency, and interactions up to third-nearest
neighbours, are needed. However at the end we will discuss our results
in the light of \cite{l:Asano}.

After these preparations we now consider the impure systems:
In Fig.~3, we present the results for the GMR obtained with
Eqn.~(\ref{e:Kubo}) and Eqn.~(\ref{e:Gmr}) for case (i), i.e.~impurities
distributed randomly across the four interfaces (see below).
 The concentrations considered
correspond to 1, 2, ..., 5 interchanged impurity pairs, e.g.~V impurities in
a Fe plane and {\it vice versa}.  Since the volume corresponds to 192 sites,
and since to every V impurity position in a Fe plane there is a neighbouring
Fe impurity in the adjacent V plane, the impurity concentrations range from
1.04\% to 5.2\%, and we have produced our random samples in such a way that
for $n_{\rm{\scriptsize{imp}}} \le 4.16$\% the impurity pairs are situated
at {\it different} interfaces, whereas for $n_{\rm{\scriptsize{imp}}} =
5.2$\% resp.~6.24\% at one of the four interfaces (resp.~two of them) two
pairs are situated. The error bars in Fig.~3 result from the
average over 10 samples, and the evaluation of Eqn.~(\ref{e:Kubo})
has been performed as in Fig.~2, but  with (2n+1)=3, separately for $s=\pm 1$. 

It is essential that in spite of the large error bars there is a clear trend
in the concentration dependence in Fig.~3, namely at first a
roughly linear {\it increase} from GMR$\sim 60$\% at
$n_{\rm{\scriptsize{imp}}}\approx 1$\% to GMR$\sim 120$\% at
$n_{\rm{\scriptsize{imp}}}\approx 4$\%, 
which is then followed by a {\it decrease} to
GMR$\sim 50$\% for $n_{\rm{\scriptsize{imp}}}\approx 5$\%.
 (Almost the same results, GMR=32\% and 53\% for the CIP and CPP cases
 respectively, are obtained with 6\% impurities.)
The small difference of the CPP-GMR with respect to the CIP-GMR is
insignificant; what is only important is that the behaviour of the GMR
with increasing concentration is the same for both cases.

The increase observed up to $n_{\rm{\scriptsize{imp}}}=4.2$\% is 
what one would expect by
an incoherent superposition of the effects from single impurities.
However, for $n_{\rm{\scriptsize{imp}}}=5.24$\%, as already mentioned,
one of the interfaces must host two impurity pairs. That this fact 
leads to a decrease of the GMR, is in agreement with the behaviour 
in case (ii), which is presented in Fig.~4: 
In this case, where -- as discussed above~-- the impurities are
concentrated at only one of the four interfaces, the GMR decreases from
$\sim 60$\% for $\approx 1$\% impurities down to GMR$\sim 0$\% at
$n_{\rm{\scriptsize{imp}}}\approx 5$\%.

The different behaviour of cases (i) and (ii) is not easily understood,
since it involves the ratio $\Gamma^{\uparrow\downarrow}/
\Gamma^{\uparrow\downarrow}$ of two conductances:
According to details of our results, which we do not present as plots, 
\cite{l:Moser}, both $\Gamma^{\uparrow\uparrow}$ and
$\Gamma^{\uparrow\downarrow}$ decrease significantly with increasing
$n_{\rm{\scriptsize{imp}}}$. Concerning $\Gamma^{\uparrow\downarrow}$, we find
that this decrease is roughly the same for the cases (i) and (ii),
respectively: E.g.~$\sigma^{\uparrow\downarrow}_{\rm{\scriptsize{CIP}}}$
 decreases from $\sim$
11$\times$10$^{-5}$ (Ohm cm)$^{-1}$ at $n_{\rm{\scriptsize{imp}}}\approx 1$\%
to $\sim 6\times 10^{-5}$ (Ohm cm)$^{-1}$ at $n_{\rm{\scriptsize{imp}}}\approx
5$\%, both for (i) and (ii), in spite of the different  spatial
impurity distributions of these cases. In contrast,
$\Gamma^{\uparrow\uparrow}$ is found to be quite {\it sensitive} to the spatial
distribution of the impurities and thus essentially responsible for the
different behaviour of the GMR: In case (i) the decay is  rather weak,
e.g.~$\sigma^{\uparrow\uparrow}_{\rm{\scriptsize{CIP}}}$ decays in a
'sub-Boltzmannian way', namely $\propto (7.67 +9.33\cdot
 n_{\rm{\scriptsize{imp}}}^{-1})$, from
17$\times 10^5$ (Ohm cm)$^{-1}$ at $n_{\rm{\scriptsize{imp}}}\approx 1$\% to
10$\times 10^5$ (Ohm cm)$^{-1}$ at $n_{\rm{\scriptsize{imp}}}\approx 4$\%, whereas
in case (ii), where the impurity pairs are randomly concentrated at one of
our four interfaces, the decay is much faster,
namely from 17$\times 10^5$ (Ohm cm)$^{-1}$ down to to $5\times 10^5$ (Ohm
cm)$^{-1}$, and essentially 'non-Boltzmannian', namely linear in
$n_{\rm{\scriptsize{imp}}}^{+1}$ instead of $n_{\rm{\scriptsize{imp}}}^{-1}$.
 This means that vertex corrections and multiple scattering could play an
essential role for $\Gamma^{\uparrow\uparrow}$ in case (ii), which is not
unreasonable in view of the essentially two-dimensional nature of the
scattering for that case, and the erratic magnetization profiles obtained in
Fig.~6 below:

\subsection{Magnetic-moment profiles}
In Fig.~5 and Fig.~6 we present in fact the distribution of magnetic moments
in disordered Fe and V interface planes with one resp.~five (non-neighbouring)
interdiffusions at the same interface. As one can see from these plots,
there is a significant
 reduction of the Fe moments in the vicinity of the V impurity. Also the V
impurities in the Fe planes have a considerable magnetic moment of $\sim
-0.8$ $\mu_B$, much higher than in the pure V interface plane, where the V
moment is only $\sim -0.25$ $\mu_B$. This agrees with first-principles
calculations of Coehoorn,
\cite{l:Coehoorn}.

\subsection{Cases (iii) and (iv)}
We get a much larger GMR effect than that obtained with 
 cases (i) and
(ii) by replacing the {\it interchange process}  of Fe and V neighbours
at the interface by the following
 particular {\it 'island substitution'} process (iii): We
substitute randomly three neighbouring V atoms  in a V interface layer
by Fe, and at a different interface in a Fe layer independently three
neighbouring Fe atoms by V, which corresponds to $\sim$ 3.12\% impurities.
 Of course there are many possibilities of such a simultaneous random
substitution of a small 'island' of three neighbouring Fe interface
atoms by V and three neighbouring V atoms at a different interface by
Fe, e.g.~atoms 1,2,3 or 2,5,6 or 2,6,10 or ... in Fig.~5a. Averaging over
10 random samples, we obtain a CIP-GMR as large as $(265\pm 120)$\% and
a CPP-GMR as large as $(300\pm 170)$\%. These  high values with large
scatter, which should be contrasted to GMR$\sim 120$\% obtained for
$n_{\rm{\scriptsize{imp}}}\approx 3$\% in case (i), are not yet
understood at present, however it is clear that the differences point
again to 'non-Boltzmannian' behaviour and the possible role of vertex
corrections. Probably it is important that the interfaces become more
''diffuse'' for 'interchange impurities' of type (i) and (ii),  whereas
in case (iii), although the width of the Fe films (V films) varies
locally, the interface remains well-defined in a sense.

Finally, in case (iv),  we discuss the situation that there is just 1 Fe
substitutional impurity (i.e.~a concentration of 0.5\%) in one of the two
{\it central} V planes of our sample. In this case, if adjacent Fe films are
magnetized in opposite direction, the Fe
 moment at the central V layer vanishes on symmetry grounds; but
if the Fe films are aligned in parallel, there is a considerable moment
induced at the Fe  impurity:

In Fig.~7 we present the magnetization profile in the relevant
central V layer for this case of mutually parallel magnetic
polarizations of the adjacent Fe films. As already mentioned for the
pure system, the V polarization in the central plane is almost
neglegibly small, but the Fe impurity moment is not: Instead, it is
magnetized antiparallel to the adjacent Fe films, with $\sim -0.5$
$\mu_B$, whereas in the Fe films themselves one has the results of
Fig.~1, namely $\mu\sim 1.5$ and 2.5 $\mu_B$ at the interfacial
resp.~central Fe layers. From this large induced negative polarization
one can imagine that the present  Fe impurity in "bulk" V induces strong
scattering effects in the case of mutually {\it parallel} polarization
of the Fe films, which is unusual, since now one expects a {\it
negative}  GMR. Averaging with respect to the few different
possibilities to place the impurity with respect to the contacts in
Eqn.~(\ref{e:Kubo}), we get in fact (up to $\sim 20$\% accuracy) the
following results: CIP-GMR$\sim$ --47\% and CPP-GMR$\sim$ --61\%.
\subsection{Discussion}
Although a direct comparison is not possible in view of the differences
discussed above, we discuss our results in the light of the paper
\cite{l:Asano} of Asano {\it et al.}. As already mentioned, these
authors treat a a very simplified s-band model only, but for larger
systems, with attached ideal leads, and with a powerful recursion
method.

 At first we stress that the values for the
CPP-GMR in \cite{l:Asano} are of the same order as ours in case (i),
namely $\sim 45$\% (compared to our $60$\%) for
$n_{\rm{\scriptsize{imp}}}\approx 1$\%. This corresponds to the
''interface roughness''
$\lambda\approx 0.06$ in \cite{l:Asano}, although these authors do not
generate the impurities by pairwise interchange, but by simple
substitutions in interface layers, which has a less drastic effect on
the local moments, see \cite{l:Pirnay}. But with increasing interface
roughness, in \cite{l:Asano} the CPP-GMR  only {\it decreases} rather
slowly, e.g.~down to $\sim 35$\% at $\lambda\approx 0.3$,
 while the CIP-GMR, which vanishes for the pure
system, {\it increases} still more slowly.
   Only, if at all sites $i$ the  potentials $V_s(i)$ are
additionally randomized by addition of terms $\delta V_s(i)$, which are
uniformly and independently distributed between $\pm W_B/2$,  Asano {\it
et al.} obtain a more drastic decrease of both  GMRs with  increasing
$W_B$, which becomes very rapid for $W_B \widetilde > 1$.

 Thus there are two  main differences to our results:
 a) Whereas in our case the CIP-GMR is only slightly lower than
the CPP-GMR (see above), in \cite{l:Asano} it always remains significantly
smaller, e.g.~by a factor $\widetilde <$ 0.25 for $\lambda\approx 0.2$.
b) Concerning the CPP-GMR, according to \cite{l:Asano}, for the pure
samples it is even somewhat higher than with impurities, whereas in our
case it practically vanishes for pure samples (see above): These
differences are due to the fact that in  \cite{l:Asano} the sample is
attached to perfect leads, i.e.~the dephasing length is $\infty$ and the
energy spectrum continuous, and the extension of the samples in x- and
y-direction is rather large, whereas in our case we assume periodic
boundary conditions, but an inelastic dephasing length as short as our
sample sizes, and similarly small widths $\Delta x$ and $\Delta y$.

Therefore, our case $n_{\rm{\scriptsize{imp}}}=0$ does not correspond to the
usual 'ballistic situation', in contrast to \cite{l:Asano}. However, for our
disordered systems the elastic scattering lengths are as short as (or even
shorter than) the inelastic dephasing length, and for the GMR of our small
and essentially cubic samples, the distinction between 'longitudinal' and
'perpendicular' may become blurred in the presence of impurities and short
inelastic dephasing.

The essential point however seems to be the following: According to \cite{l:Asano}, disorder and impurities apparently
always {\it reduce} the CPP-GMR, whereas for our small systems with the
strong 'magnetic contrast' of Fe and V and the presence of strongly
different local situations, also an {\it increase} can happen, and even
a {\it negative} GMR may be possible. Actually however, in case (i) and
(ii), if there are two or more impurities at an interface plane,
i.e.~for $n_{\rm{\scriptsize{imp}}}\ge 1.04$\% in case (ii), or  for
$n_{\rm{\scriptsize{imp}}}\ge 5.2$\% in case (i), also in our results
the GMR always {\it decreases} with increasing disorder.

 Experimentally it has been found in \cite{l:Kelly} that increasing the
interface roughness by Xe$^+$ irradiation leads to a significant
enhancement of the GMR, however beyond a certain irradiation dose the
GMR decreased again, which would fit to our main scenario (i).

\section{Conclusions} We have applied a self-consistent semi-empirical
{\it almost ab-initio} approach to small samples, to calculate the
magnetic properties and the CIP- and CPP-GMR effects for
(001)-(3--Fe/3--V)$_\infty$ multilayers, with impurities generated by
randomly interchanging neighbouring Fe and V atoms (i) at all four
interfaces, and (ii) only one of the interfaces, for impurity
concentrations ranging from 1\% to 5\% (sometimes 6\%). In case (i) we
observed an increase of the GMR from $\sim 60$\% for 1\% to $\sim 120$\%
at 4\% impurities, followed by a decrease back to $\sim 50$\% for 5 and
6\% impurities, whereas in case (ii), we have found a decrease of the
GMR from $\sim 60$\% for 1\% to $\sim 0$ at 5\% impurities. Still much
larger GMR values of $\sim 250$ to $\sim 300$\% have been obtained in
case (iii) for impure systems with small "islands" of three randomly
chosen neighbouring  Fe substitutions in a V interface layer and
simultaneous, but independent V substitutions in a Fe layer at a
different interface, which corresponds to $\approx$ 3\% impurities.
Finally, in case (iv), we considered  Fe impurities in "bulk" V,  by
replacing one of  16 V atoms in one of two central V layers by Fe. This
leads in our calculation to a negative GMR of $\sim$ (-50\%). It is
remarkable that here a negative (="inverse") GMR has been obtained in a
binary system, whereas hitherto this was observed only in ternary
systems \cite{l:Mertig2}.

Of course our results should be taken with care: One should be aware of the 
smallness of the systems (192 atoms), the small 'dephasing distances'
 between the voltage contacts ($\Delta x=\Delta y\approx 12$ $\AA$, $\Delta
z\approx 18$ $\AA$), the large number of strongly reflecting
Fe/V-interfaces (two of three monolayers are interface layers), the small
 concentrations  of strongly scattering impurities involved (up to 5\% only,
Fe impurities in V and {\it vice versa}), and the large error bars of the
sample averages. However in spite of these {\it caveats}, the results seem
to justify the statement that the influence of magnetic impurity scattering
on the GMR may be more complicated than expected.
\subsection*{Acknowledgements}
 The authors would like to thank S.~Krompiewski, F.~S\"uss and M.~B\"ohm
for valuable discussions. The computations have been performed at the
computing centers of the university at Regensburg, the LRZ in Munich and
the HLRZ in J\"ulich.

\newpage  
\baselineskip17pt
{ 
 {\centerline{\bf Figure Captions}}

{\bf Fig.~1} Comparison of the magnetic moments of the pure
(001)-(3--Fe/3--V)$_\infty$ multilayers with a) mutually parallel
resp.~b) antiparallel magnetization of the Fe moments, obtained with the
{\it ab-initio} LMTO program (empty squares) and our self-consistent
semi-empirical LCAO approach (full squares). Our elementary cell
consists of altogether 12 planes with 16 atoms each, and periodic
boundary conditions in x-,y-, and z-direction.

{\bf Fig.~2}: Conductivities of the pure (001)-(3--Fe/3--V)$_\infty$
multilayers, with contacts as described in the text. "Ferromagnetic"
resp.~"antiferromagnetic" means that the magnetization directions of the
Fe layers are mutually parallel resp.~antiparallel. The number (2n+1)
of contributing states for given $s$, in the vicinity of the
spin-dependent Fermi-energy $E_f(s)$, ranges from 3 to 11. The dotted
lines represent the averages of the five cases of (2n+1). The  apparent
difference beween the CPP and CIP  conductivities ( "current
perpendicular to the planes" and "current in the planes") is
not realistic: It results simply from the different distances of the
contacts for CIP and CPP, and can be scaled away (see the text). Finally,
within our accuracy, the GMR is zero for the present pure system,
i.e.~the results for the different mutual magnetizations are practically
identical. For further details see the text.

{\bf Fig.~3}:  CIP- and CPP-GMR (see Equ.~(\ref{e:Gmr})) for impurity
distributions of class (i). This class corresponds to "interchange 
impurities" concentrated at up to four {\it different} interfaces, n
with 16 sites per plane. For concentrations of up to 4.16\%, every (Fe-V)
pair of interchanged atoms has its own interface, whereas in case of 5.2\%
impurities, at one of four interfaces there are not one, but two of such
(Fe-V) pairs. Note that for this class the GMR is mainly {\it
increasing} with the impurity concentration.

{\bf Fig.~4}: The same as Fig.~3, however for impurity distributions of
class (ii). This class corresponds to "interchange 
impurities" concentrated at only one of four interfaces with 16 sites
per plane; i.e.~for (Fe-V)-pairs at the "impure interface", the Fe-atoms
 are substituted by V and {\it vice versa}. Note that for this class the
 GMR is decreasing with increasing impurity concentration.

{\bf Fig.~5}:  In this figure, the spatial distribution of the magnetic
moments is presented for a Fe interface plane with one V impurity
(Fig.~5b) and for a V interface plane with one Fe impurity (Fig.~5c).
The relation between the {\it numbers} 1,2,...,16 given to the atoms and
their {\it positions} in space follows from Fig.~5a.

{\bf Fig.~6}: The same as in Fig.~5, however for interface planes with 5
impurities (out of 16 sites).

{\bf Fig.~7}: The same as in Fig.~5, but for case (iv), i.e.~there is
now only one Fe impurity in a {\it central} V plane. The Fe films are
magnetized in parallel ("ferromagnetic coupling"). In this case there
results a negative GMR, see the text.
 

\newpage

\input epsf
\begin{figure}[htb]
\epsfysize=16cm
\epsfbox{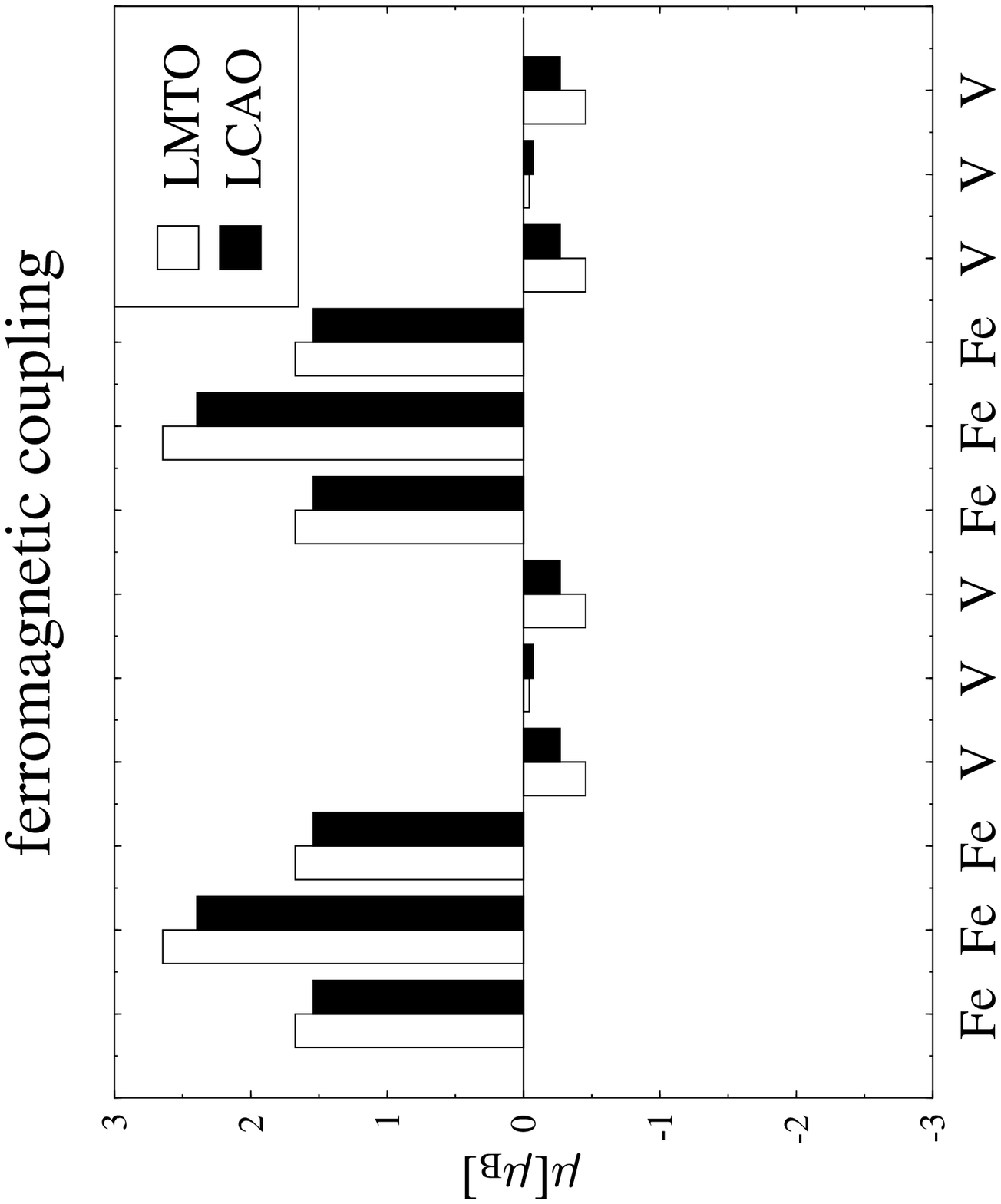}
{\centerline{\bf Fig.1a}
}
\end{figure}
\begin{figure}[htb]
\epsfysize=16cm
\epsfbox{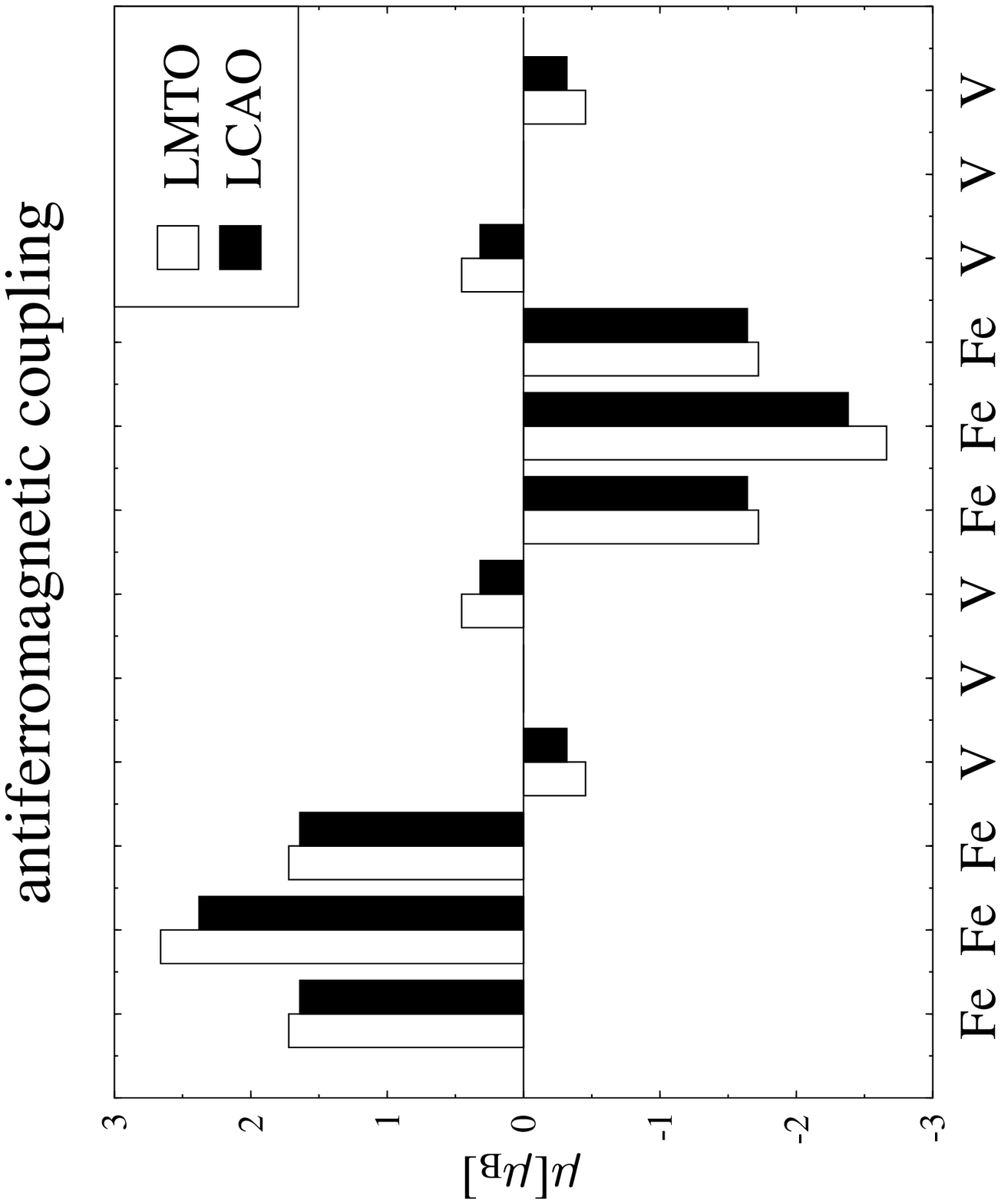}
{\centerline{\bf Fig.1b}
}
\end{figure}
\begin{figure}[htb]
\epsfysize=16cm
\epsfbox{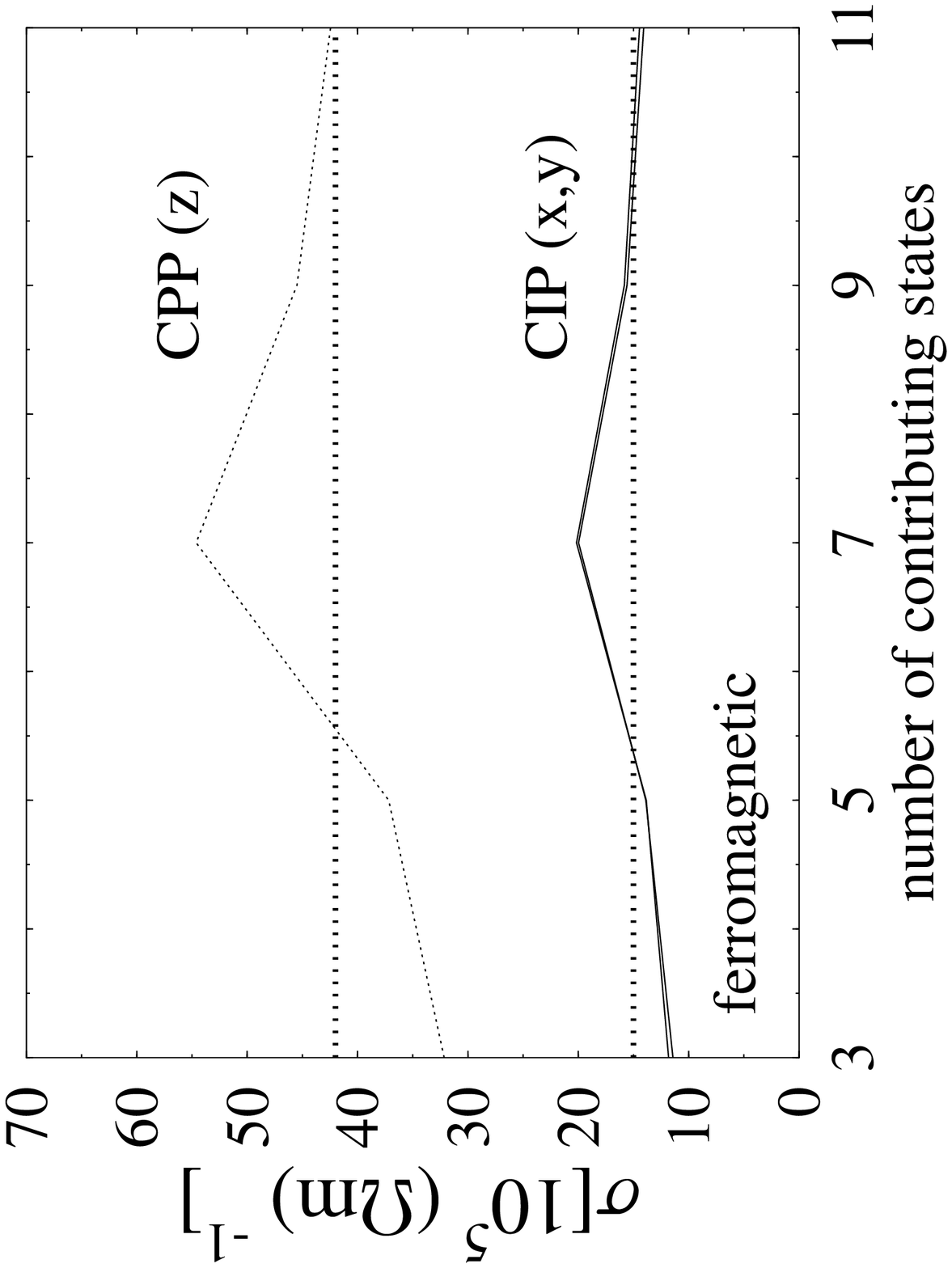}
{\centerline{\bf Fig.2a}
}
\end{figure}
\begin{figure}[htb]
\epsfysize=16cm
\epsfbox{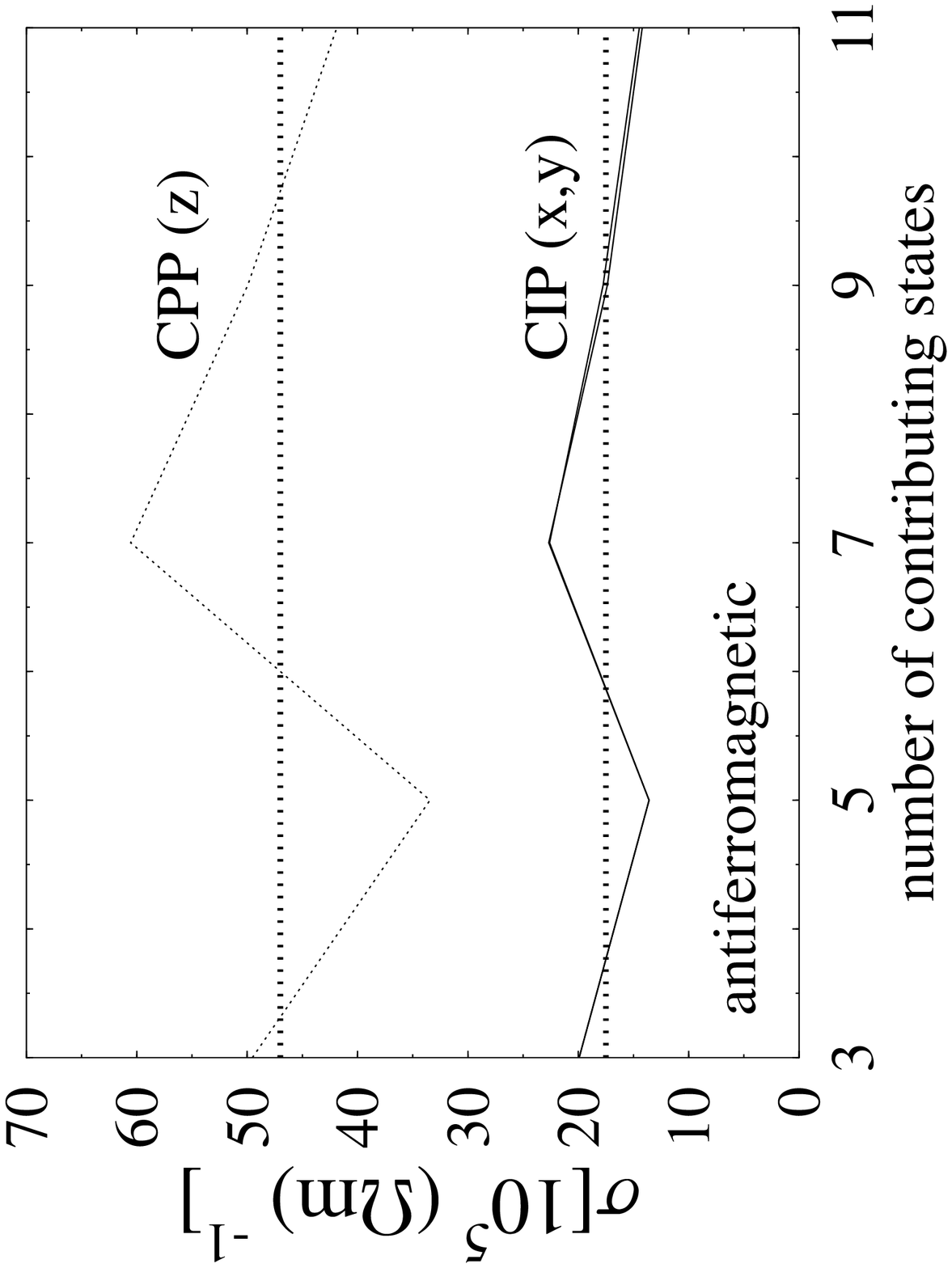}
{\centerline{\bf Fig.2b}
}
\end{figure}
\begin{figure}[htb]
\epsfysize=16cm
\epsfbox{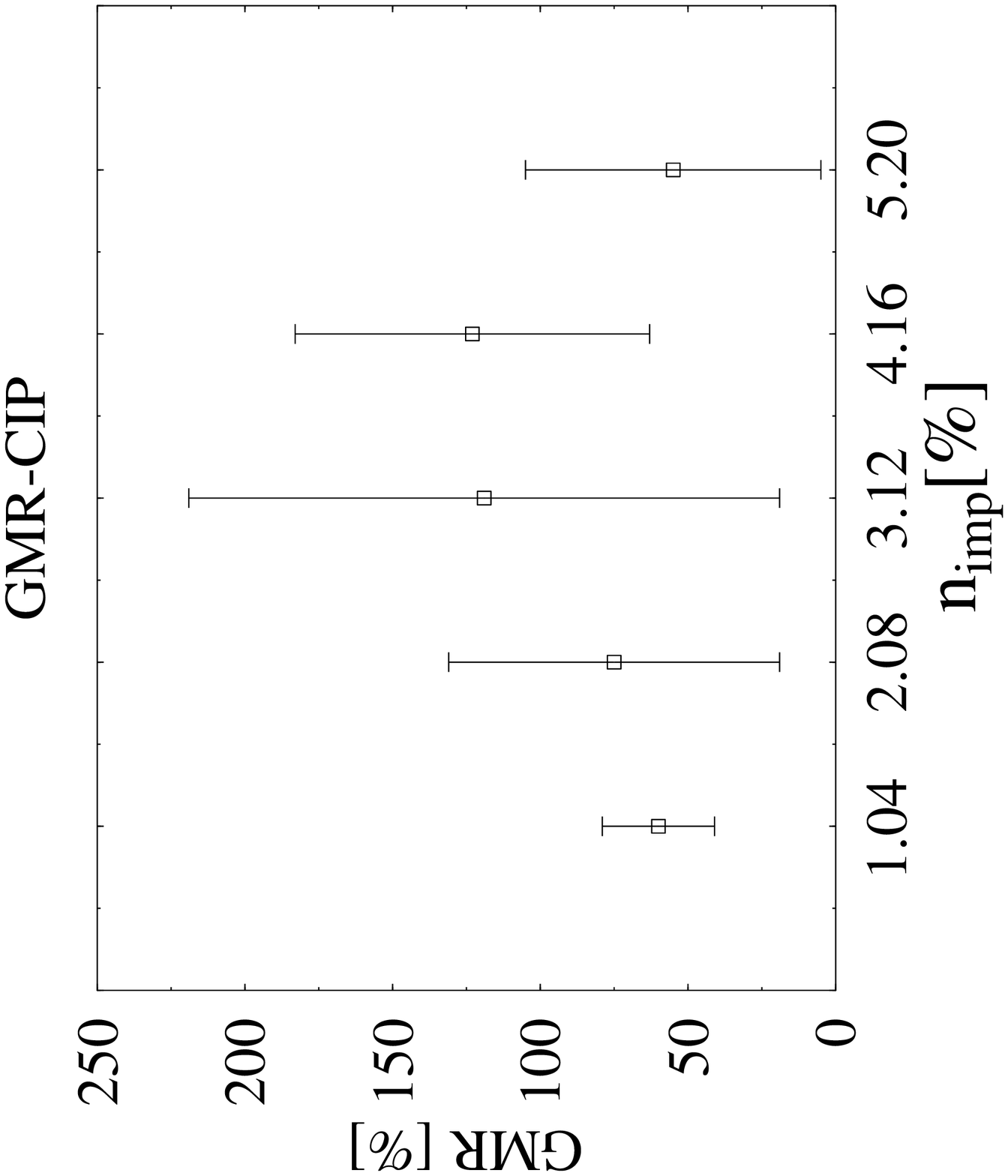}
{\centerline{\bf Fig.3a}
}
\end{figure}
\begin{figure}[htb]
\epsfysize=16cm
\epsfbox{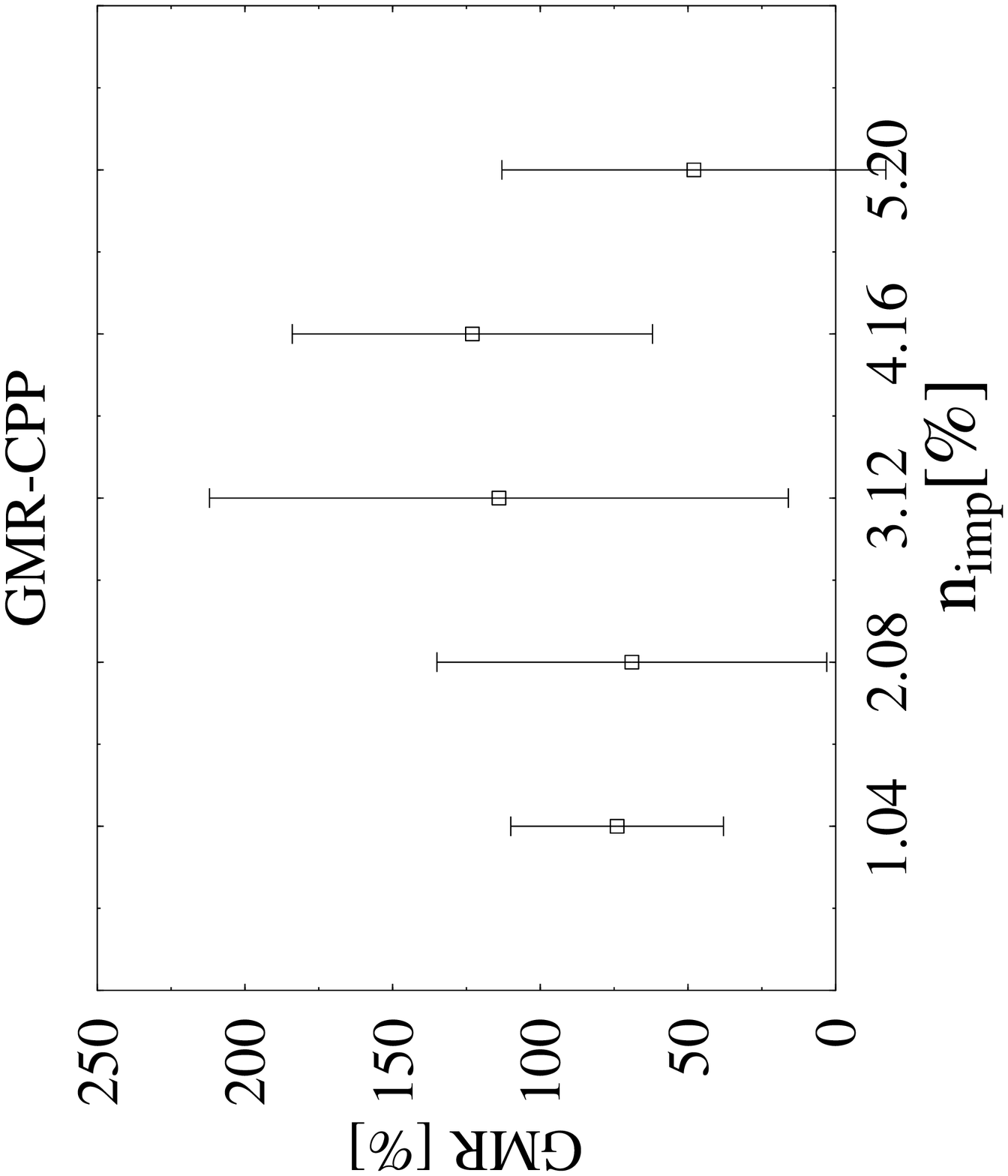}
{\centerline{\bf Fig.3b}
}
\end{figure}
\begin{figure}[htb]
\epsfysize=16cm
\epsfbox{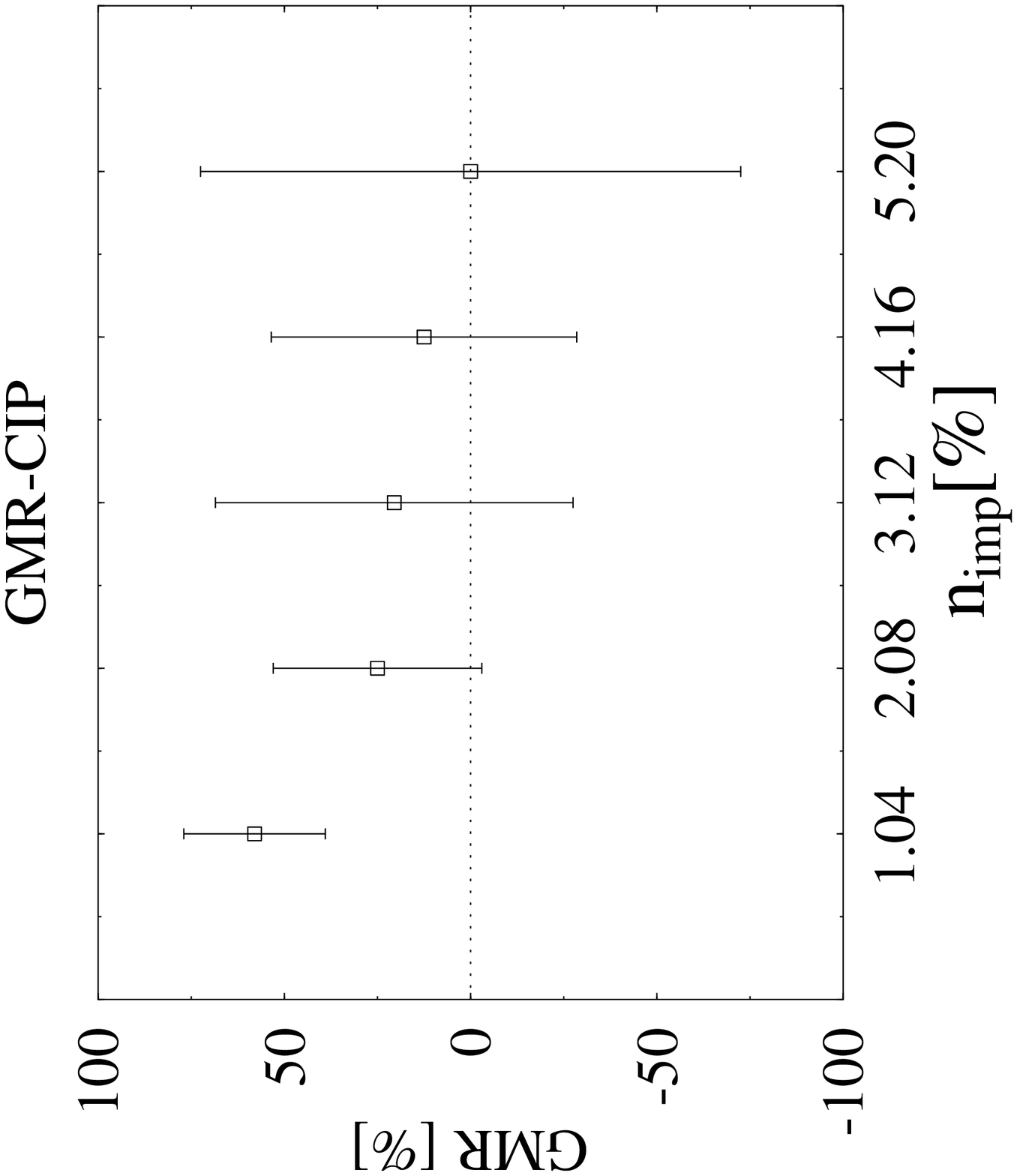}
{\centerline{\bf Fig.4a}
}
\end{figure}
\begin{figure}[htb]
\epsfysize=16cm
\epsfbox{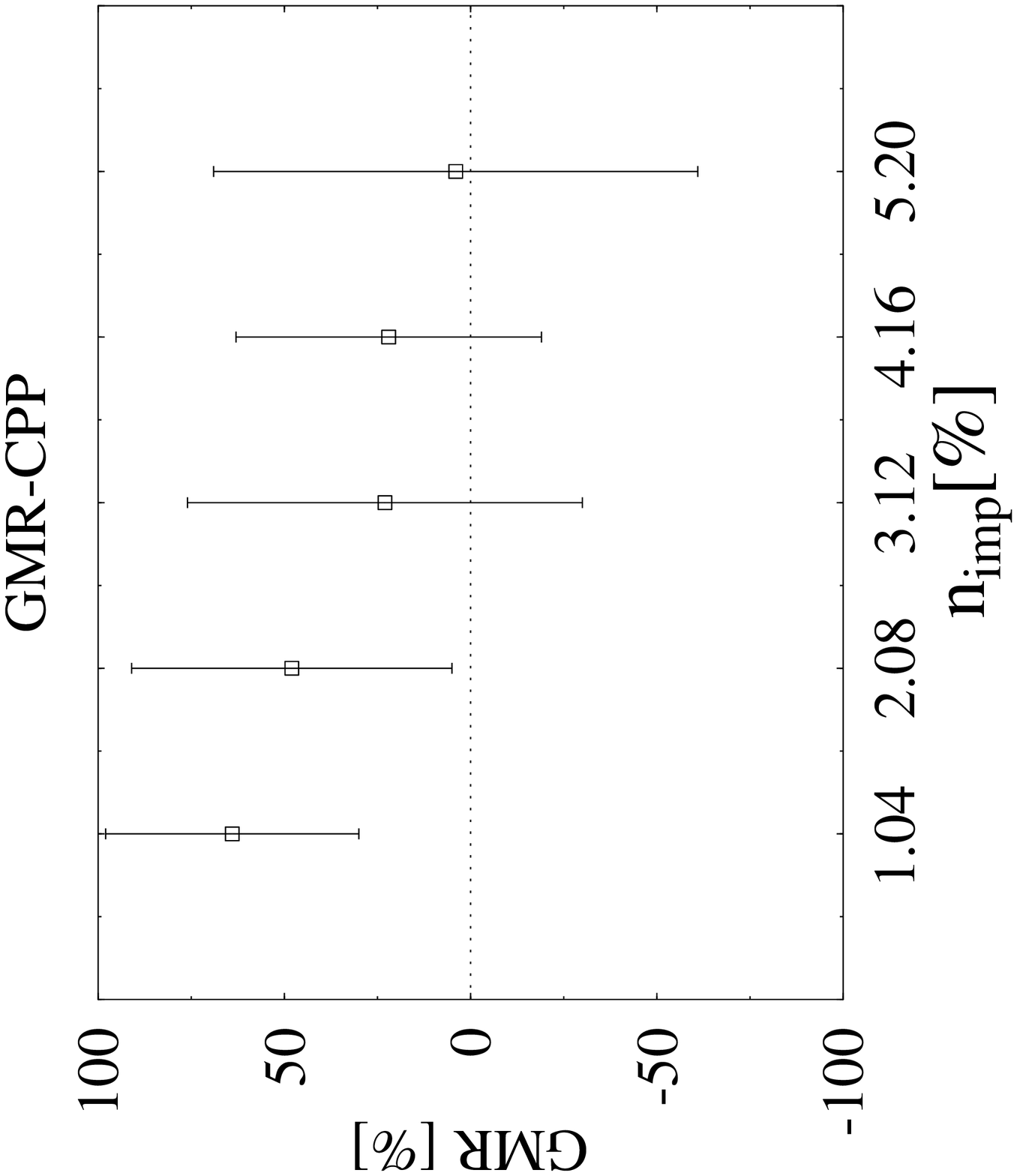}
{\centerline{\bf Fig.4b}
}
\end{figure}
\begin{figure}[htb]
\epsfysize=16cm
\epsfbox{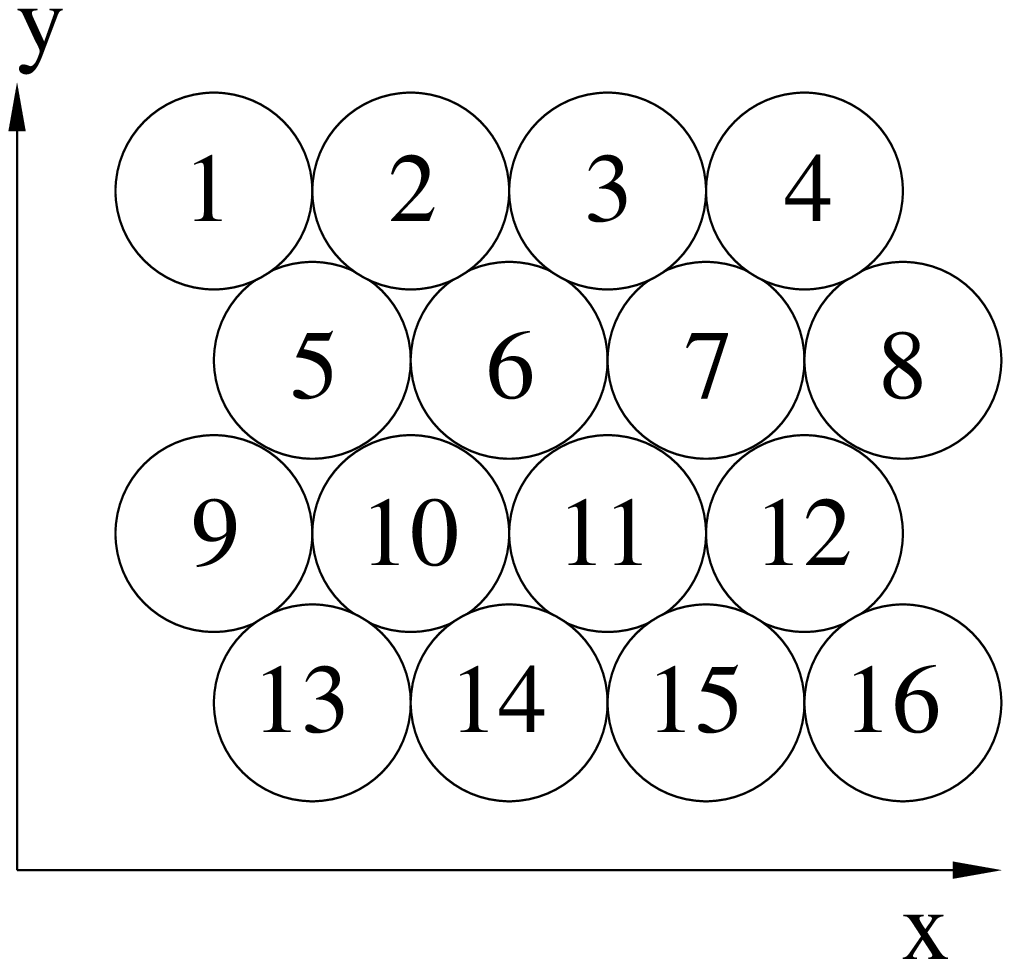}
{\centerline{\bf Fig.5a}
}
\end{figure}
\begin{figure}[htb]
\epsfysize=16cm
\epsfbox{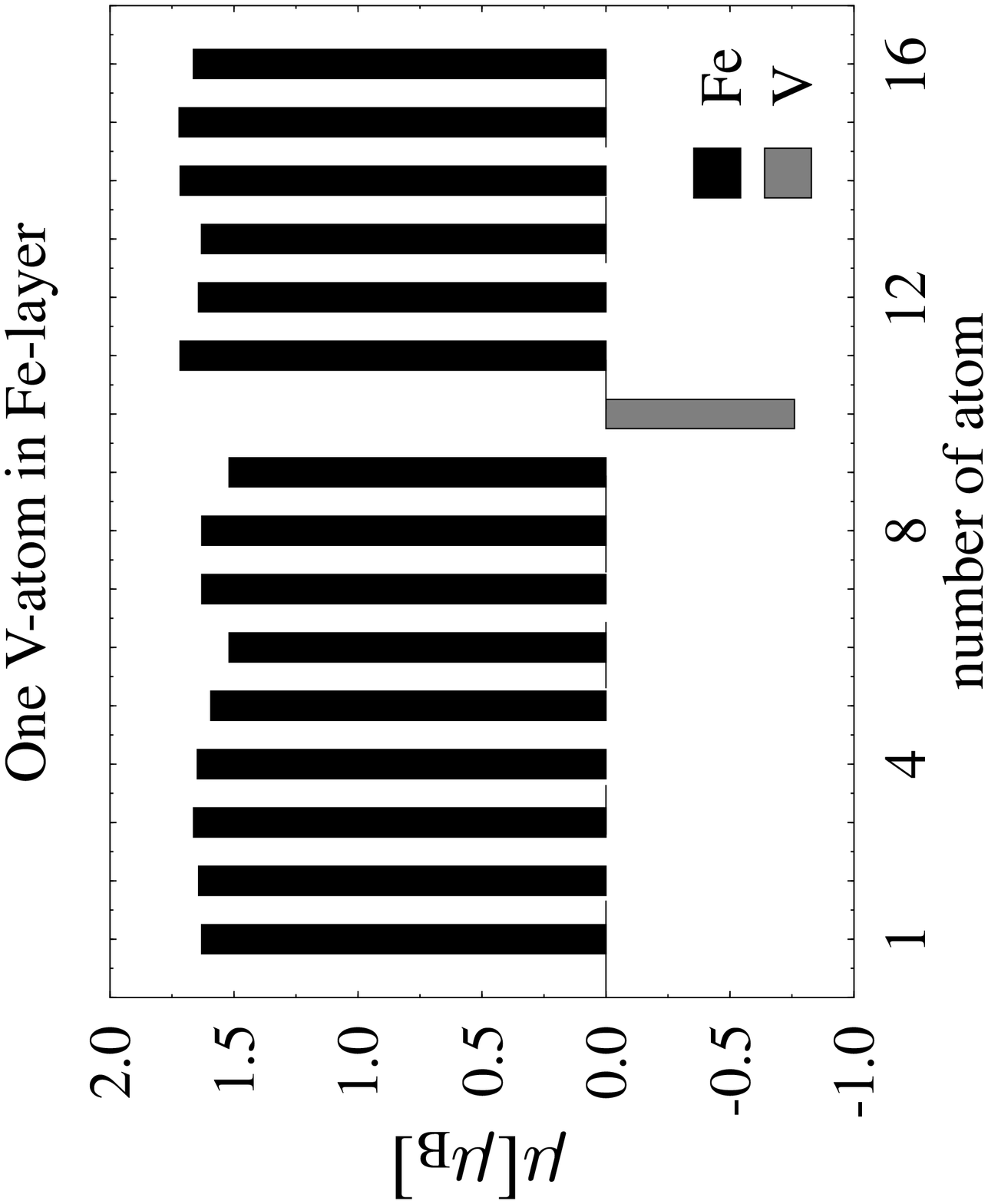}
{\centerline{\bf Fig.5b}
}
\end{figure}
\begin{figure}[htb]
\epsfysize=16cm
\epsfbox{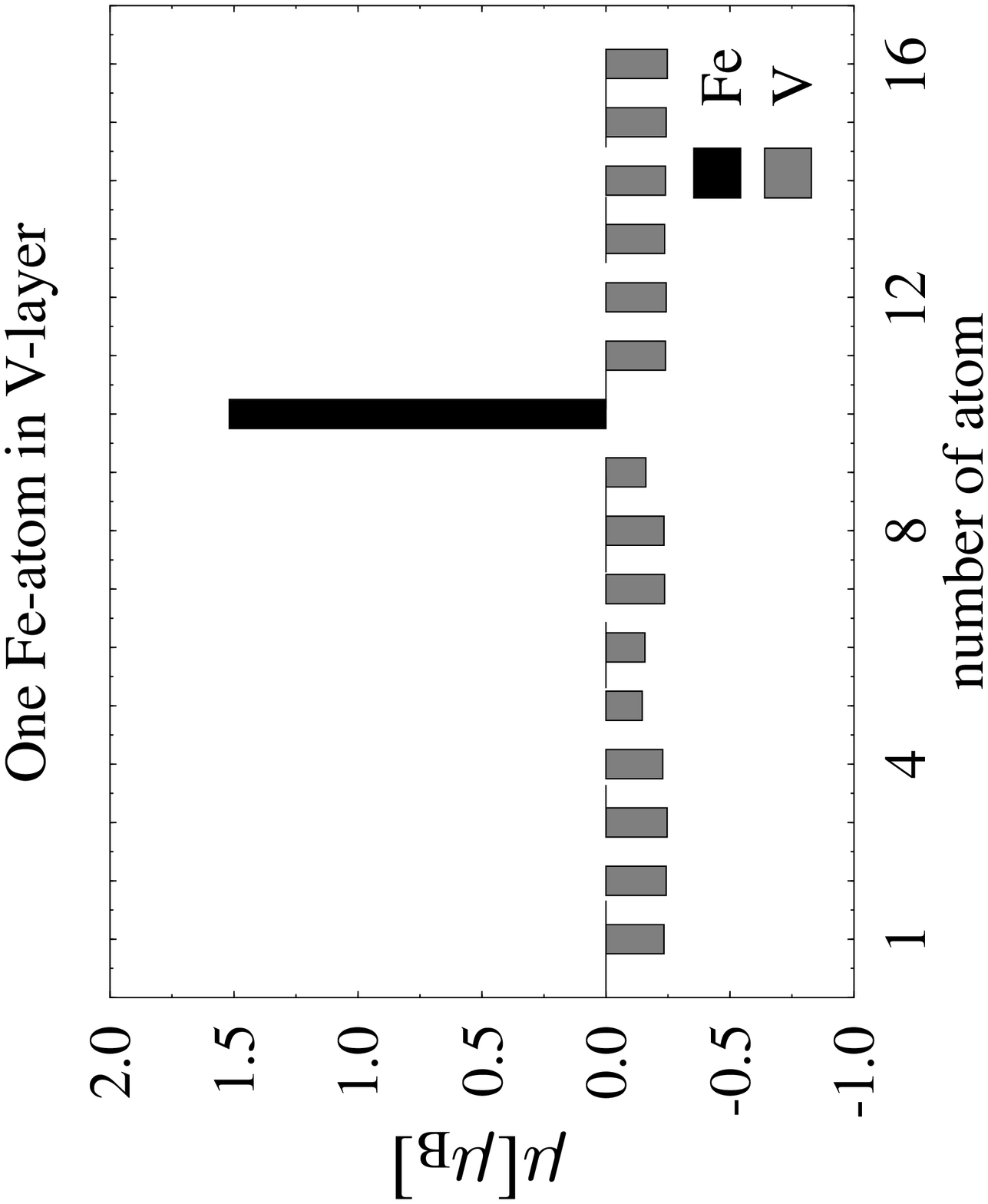}
{\centerline{\bf Fig.5c}
}
\end{figure}
\begin{figure}[htb]
\epsfysize=16cm
\epsfbox{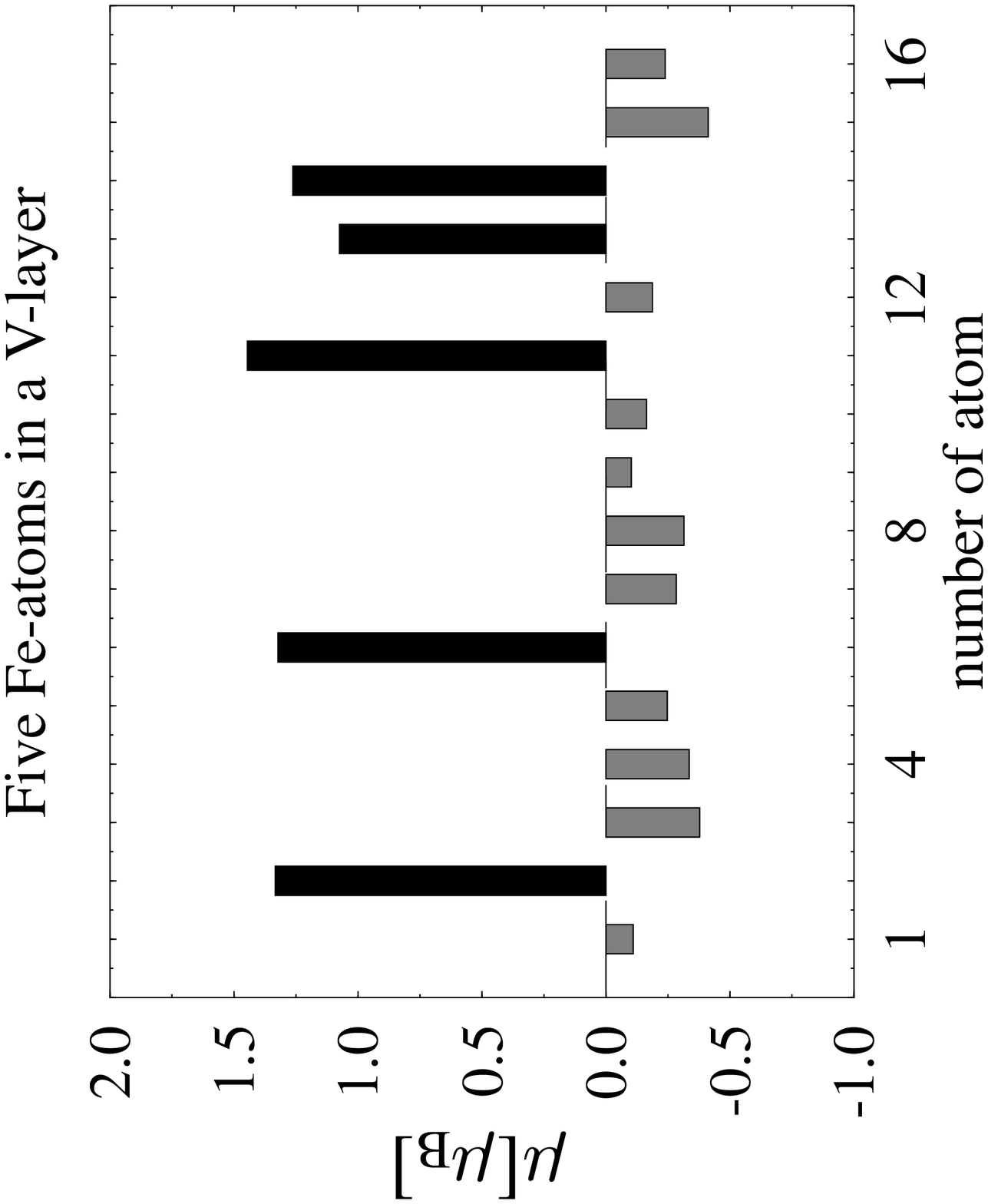}
{\centerline{\bf Fig.6a}
}
\end{figure}
\begin{figure}[htb]
\epsfysize=16cm
\epsfbox{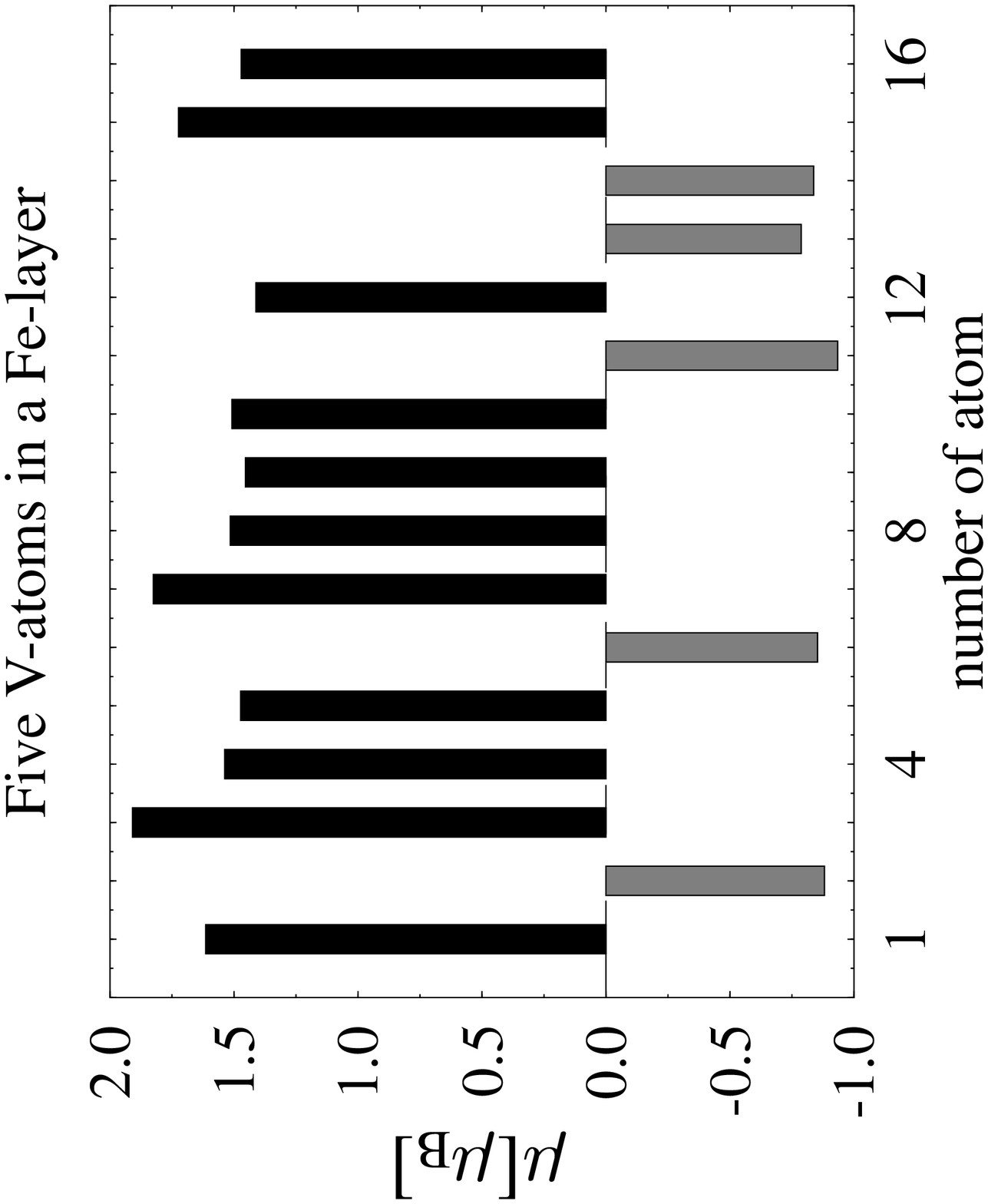}
{\centerline{\bf Fig.6b}
}
\end{figure}
\begin{figure}[htb]
\epsfysize=16cm
\epsfbox{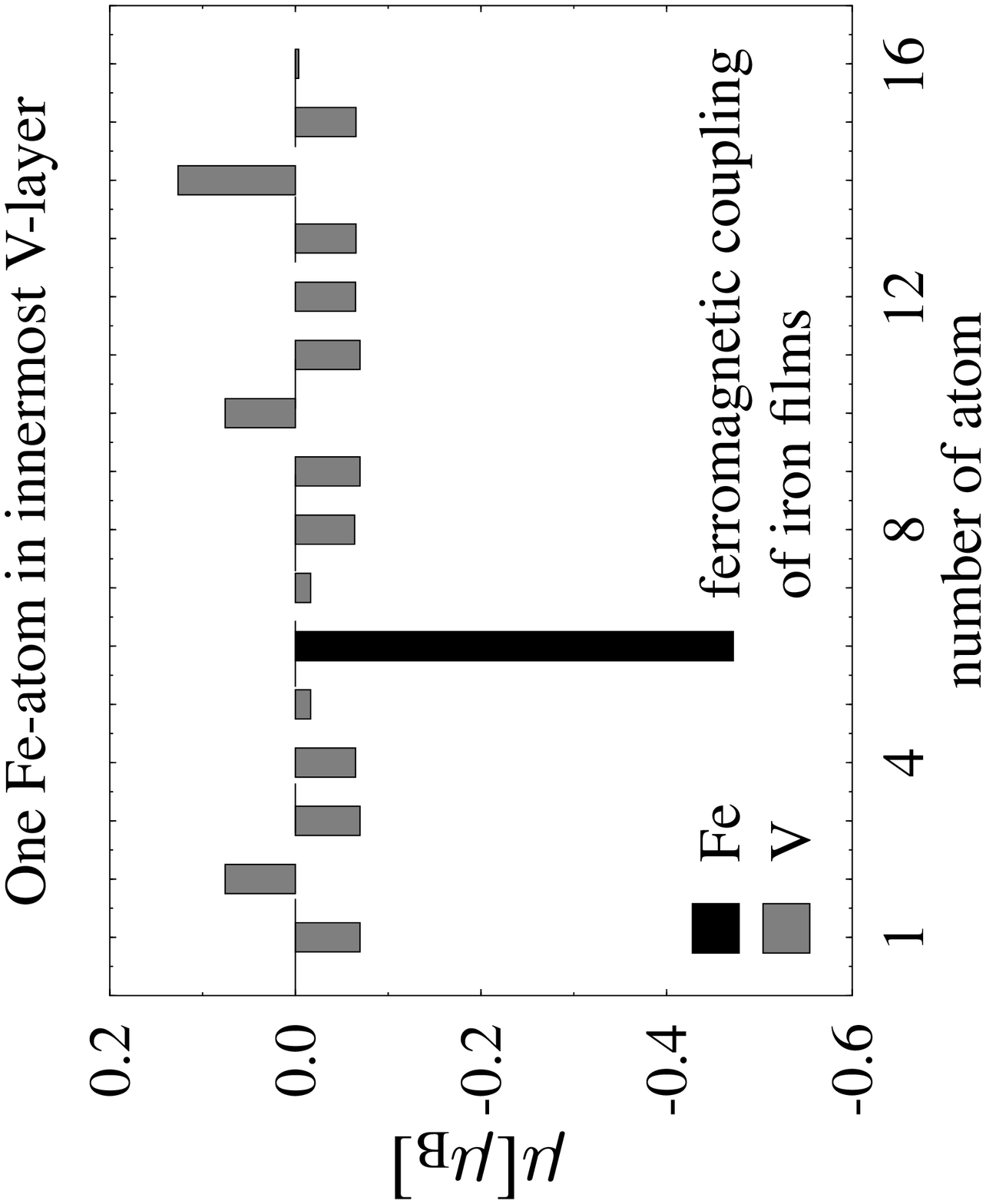}
{\centerline{\bf Fig.7}
}
\end{figure}
\end{document}